\documentclass{elsart}

\usepackage{graphicx}
\usepackage{amssymb}

\begin{document}

\begin{frontmatter}

% Title, authors and addresses

% use the thanksref command within \title, \author or \address for footnotes;
% use the corauthref command within \author for corresponding author footnotes;
% use the ead command for the email address,
% and the form \ead[url] for the home page:
% \title{Nuclear expansion and symmetry energy of hot nuclei}
% \thanks[label1]{}
% \author{D.V. Shetty\corauthref{}\thanksref{label2}}
 \author{}
 \ead{shetty@comp.tamu.edu}
% \ead[url]{home page}
% \thanks[label2]{}
 \corauth[cor1]{Corresponding author}
 
% \address{Address\thanksref{label3}}
% \thanks[label3]{}

\title{Nuclear expansion and symmetry energy of hot nuclei}

% use optional labels to link authors explicitly to addresses:
% \author[label1]{S.J. Yennello, G.A. Souliotis}
% \address[label1]{Cyclotron Institute}
% \address[label2]{}

\author{D.V. Shetty*,}
\author{G.A. Souliotis,}
\author{S. Galanopoulos,}
\author{Z. Kohley,}
\author{S.N. Soisson,}
\author{B.C. Stein,}
\author{S. Wuenschel,}
\author{and S.J. Yennello,}

\address{Cyclotron Institute, Texas A$\&$M University, College Station, TX 77843, USA}

\begin{abstract}
The decrease in the symmetry energy of hot nuclei populated in $^{58}$Ni + $^{58}$Ni, $^{58}$Fe + $^{58}$Ni and $^{58}$Fe + $^{58}$Fe 
reactions at beam energies of 30, 40 and 47 MeV/nucleon, as a function of excitation energy is studied. It is observed that 
this decrease is mainly a consequence of increasing expansion or decreasing density rather than the increasing temperature. The 
results are in good agreement with the recently reported microscopic calculation based on the Thomas-Fermi approach. An empirical relation 
to study the symmetry energy of finite nuclei in various mass region is proposed. 
\end{abstract}

\end{frontmatter}

% main text

The symmetry energy, which is the difference in energy per nucleon between the pure neutron matter and the symmetric nuclear 
matter, is a topic of significant interest \cite{STO07}. Traditionally, the symmetry energy coefficient of nuclei has been 
extracted by fitting the binding energy in their ground state with various versions of the liquid drop mass formula \cite{LUN03}. 
The properties of nuclear matter are then determined by theoretically extrapolating the nuclear models designed to study the 
structure of real nuclei. However, real nuclei are cold ($T$ $\approx$ 0 MeV), nearly symmetric ($N$ $\approx$ $Z$) and found at 
equilibrium density ($\rho_{o}$ $\approx$ 0.16 fm$^{-3}$). It is not known how the symmetry energy evolves at temperatures and 
densities away from normal nuclear conditions. In particular, the density and the energy dependence of the symmetry energy are 
actively being sought. Information on the symmetry energy as a function of density and temperature is crucial for many 
astrophysical calculations such as, determining the structure and cooling of neutron stars, and simulating the dynamics of 
supernova collapse \cite{STE05}. It is also important in studies related to the structure of neutron-rich nuclei, where it 
is known to be intimately related to the neutron skin thickness \cite{BRO00}.

The disassembly of a hot nucleus into several light and heavy fragments in a process called multifragmentation
\cite{BON95,BON85,BOT02} provides an important means of studying nuclei away from normal nuclear conditions. Over the last 
several decades many measurements have been carried out. Some of the important results that have emerged from these studies 
are: i) The temperature of the hot nucleus increases rapidly with increasing excitation energy, until a near flattening or 
a plateau-like region appears at higher excitation energy where the temperature remains fairly constant ({\it {caloric curve}}) 
\cite{NAT02}. ii) The density of the hot nucleus decreases as the excitation energy increases \cite{NATO02,VIO04} 
due to thermal expansion. iii) The symmetry energy, obtained from the yield distribution of the fragments following the 
disassembly of hot nuclei, is significantly lower than those normally assumed in various model calculations \cite{SHE04,IGL06,FEV05,HEN05}. 

Currently, there exist no detailed understanding of how the symmetry energy evolves with the excitation energy of hot nuclei. 
In this work, we examine the excitation energy dependence of the symmetry energy in multifragmentation of hot nuclei 
($A$ $\sim$ 100) populated in $^{58}$Ni + $^{58}$Ni, $^{58}$Fe + $^{58}$Ni and $^{58}$Fe + $^{58}$Fe reactions, and propose 
an empirical relation for studying the symmetry energy in various mass regions.

It is known that in multifragmentation reaction, the ratio of fragment isotopic yields from two different reactions, 1 and 2, 
$R_{21}(N,Z)$, follows an exponential dependence on the fragment neutron number ($N$) and the proton number ($Z$); an 
observation known as isoscaling \cite{BOT02,TSAN01,TSANG01}. The dependence is characterized by a simple relation given as,

\begin{equation} 
            R_{21}(N,Z) = Y_{2}(N,Z)/Y_{1}(N,Z) = C.\exp(\alpha N + \beta Z),
\end{equation}

where $Y_{2}$ and $Y_{1}$ are the fragment yields from the neutron-rich and neutron-deficient systems, respectively. $C$ is 
an overall normalization factor, and the quantities $\alpha$ and $\beta$ are the isoscaling parameters. 

For the present study, we make use of the experimentally determined fragment yield distribution and the isoscaling parameter
$\alpha$, obtained from the above scaling relation, for two different pairs of reactions, $^{58}$Fe + $^{58}$Ni and 
$^{58}$Ni + $^{58}$Ni, and $^{58}$Fe + $^{58}$Fe and $^{58}$Ni + $^{58}$Ni, at beam energies of 30, 40 and 47 MeV/nucleon. The 
details of the measurements and the extraction of isoscaling parameter can be found in Ref. \cite{SHE04}. 

Fig. 1(a) shows the experimentally obtained isoscaling parameter $\alpha$ (symbols), as a function of excitation energy for the 
Fe + Fe and Ni + Ni pair (inverted triangles), and the Fe + Fe and Fe + Ni pair (solid circles), of reactions. The excitation 
energy of the multifragmenting source for each beam energy was determined by simulating the initial stage of the collision 
dynamics using the Boltzmann-Nordheim-Vlasov (BNV) model calculation \cite{BAR02}.  The results were obtained at a time 
around 40 - 50 fm/c after the projectile had fused with the target nuclei and the quadrupole moment of the nucleon coordinates 
(used for identification of the deformation of the system) approached zero. These excitation energies were compared with those 
obtained from the systematic calorimetric measurements (see Ref. \cite{NAT02}) for systems with mass ($A$ $\sim$ 100) similar 
to those studied in the present work, and were found to be in good agreement with each other. One observes from the figure that 
there is a systematic decrease in the absolute value of the isoscaling parameter with increasing excitation energy for both pair 
of reactions. In addition, the $\alpha$ parameter for the $^{58}$Fe + $^{58}$Fe and $^{58}$Ni + $^{58}$Ni is about twice as 
large compared to the one for the $^{58}$Fe + $^{58}$Ni and $^{58}$Ni + $^{58}$Ni pair of reactions. 

Fig. 1(a) also shows a 
comparison between the Statistical Multifragmentation Model (SMM) \cite{BON95,BOT01}  predicted isoscaling parameter (curves), 
and the experimentally determined isoscaling parameter for the two pairs of systems. The dashed curves in the figure correspond to 
the SMM calculated isoscaling parameter for the primary fragments, and the solid curves to the same for the secondary fragments. 
The width in the curve is the measure of the uncertainty in the inputs to the calculation. The initial parameters such as, the 
mass, charge and excitation energy of the fragmenting source for the SMM calculation, were obtained from the BNV calculations 
as discussed above. To account for the possible uncertainties in the source parameters due to the loss of nucleons during 
pre-equilibrium emission, the calculations were also carried out for smaller source sizes. The break-up density in the 
calculation was taken to be multiplicity-dependent and was varied from approximately 1/2 to 1/3 the saturation density. This 
was achieved by varying the free volume with the excitation energy as described in Ref. \cite{BON95}. The form of the variation 
adopted was taken from the work of Bondorf {\it {et al.,}} \cite{BON85,BON98} (and shown by the solid curve in Fig. 1(d)). From 
the above comparison, one observes that the experimentally observed decrease in the $\alpha$ with increasing excitation energy 
and decreasing isospin difference $\Delta(Z/A)^2$ of the systems, is reproduced reasonably well by the SMM calculation. 

In Fig. 1(b), is shown the temperature as a function of excitation energy ({\it {caloric curve}}) obtained from the above SMM 
calculation.  These are shown by the solid circle and inverted triangle symbols. Also shown in the figure are the experimentally 
measured caloric curve data compiled by Natowitz {\it {et al.}} \cite{NAT02} from various measurements for the mass range 
studied in this work. The data from these measurements are shown collectively by solid star symbols and no distinction is made 
among them. It is evident from the figure that the temperatures obtained from the SMM calculation are in good agreement with 
the overall experimental trend in the caloric curve. By allowing the break-up density to evolve with the excitation energy, a 
near plateau that agrees with the experimentally measured caloric curves is thereby obtained. 

The symmetry energy in the statistical model calculations is related to the isoscaling parameter through the 
relation \cite{BOT02,TSAN01,TSANG01},

\begin{equation}
      \alpha ^{prim} = \frac{4C_{sym}}{T} {[(Z/A)_{1}^{2} - (Z/A)_{2}^{2}]}
\end{equation}

where $\alpha^{prim}$, is the isoscaling parameter for the primary (hot) fragments , {\it {i.e.}}, before they sequentially 
decay into secondary (cold) fragments.  $Z_1$, $A_1$ and $Z_2$, $A_2$ are the charge and the mass numbers of the composite 
systems from reactions 1 and 2, respectively. $T$ is the common temperature of the systems and $C_{sym}$ is the symmetry 
energy. In the above equation, the entropic contribution to the symmetry free energy is assumed to be small (the contribution 
becomes important at densities below 0.008 $fm^{-3}$ \cite{HOR05}), the symmetry energy can therefore be substituted for the 
free energy. The symmetry energy in the calculation was varied until a reasonable agreement between the calculated $\alpha$ 
for the secondary fragments and the measured $\alpha$, as shown in Fig. 1(a), was obtained. 

The symmetry energy thus obtained is shown in Fig. 1(c) as a function of excitation energy. A steady decrease in the symmetry 
energy with increasing excitation energy is observed for both pairs of systems. The effect of the symmetry energy evolving 
during the sequential de-excitation of the primary fragments \cite{IGL06} was also estimated, and these are reflected in the 
large error bars shown in Fig. 1(c).

In a recent schematic calculation by Sobotka {\it {et al.}} \cite{SOB04}, and a fully microscopic calculation by 
De {\it {et al}} \cite{DE06}, it has been shown that the plateau in the caloric curve could be a consequence of the thermal 
expansion of the system at higher excitation energy and decreasing density. By assuming that the decrease in the break-up 
density, as taken in the present statistical multifragmentation calculation, can be approximated by the expanding Fermi gas 
model, and the temperature in Eq. 2 and the temperature in the Fermi-gas relation are related, one can extract the density as 
a function of excitation energy using the simple relation,

\begin{equation}
      T = \sqrt {K_{o} (\rho / \rho_o)^{2/3} E^*}
\end{equation}

In the above expression, the momentum and the frequency dependent factors in the effective mass ratio are assumed to be one, 
as is to be expected for high excitation energies and low densities studied in this work \cite{HAS86,SHL9091}.

Using the temperatures obtained from the SMM calculation and assuming $K_{o}$ = 10 in Eq. 3, the densities obtained as a function 
of excitation energy for the two pairs of systems are shown in Fig. 1(d) (solid circles and inverted triangle symbols). For 
comparison, we also show the break-up densities obtained from the analysis of the apparent level density parameters required 
to fit the measured caloric curve by Natowitz {\it {et al.}} \cite{NATO02}.  One observes that the present results obtained by 
requiring to fit the measured isoscaling parameters and the caloric curve are in good agreement with those obtained by 
Natowitz {\it {et al}}. To verify the validity of Eq. 3,  the caloric curve obtained using the above densities and excitation 
energies, is shown by the dotted curve in Fig. 1(b). The small discrepancy between the dashed curve and the data (solid stars) 
below 4 MeV/nucleon is due to the approximate nature of the Eq. 3 being used.   

From figures, 1(a), 1(b), 1(c) and 1(d), one observes that the decrease in the experimental isoscaling parameter $\alpha$, the 
flattening of the temperature, the decrease in the symmetry energy and the  break-up density, with increasing excitation energy 
are all correlated. It can thus be concluded that the expansion of the system with excitation energy during the multifragmentation 
process, leads to a decrease in the isoscaling parameter, symmetry energy, density, and the flattening of the temperature. 

Since the temperature in the present work remains nearly constant for the range of excitation energy studied, the  observed 
decrease in the symmetry energy with increasing excitation energy must be a consequence of decreasing density, rather than the 
increasing temperature.

The decrease in the symmetry energy with increasing excitation energy observed is in close agreement with 
the recently reported calculation of Samaddar {\it {et al}} \cite{SAM07}. This microscopic calculation  is based on the 
Thomas-Fermi formulation that accounts for thermal and expansion effects in finite nuclei. Fig. 2 (second panel from the top) 
shows a comparison between the symmetry energy obtained from the present study and those from the Thomas-Fermi calculation. The 
solid circle and inverted triangle symbols correspond to the symmetry energy obtained from the present study. The solid squares 
correspond to the data measured in a previous study \cite{SOU0607} at lower excitation energies. The solid blue curve is the 
Thomas-Fermi calculation. One observes a reasonably good agreement between the experimentally determined and the theoretically 
calculated symmetry energy over a broad range of excitation energy. 

As mentioned earlier, the symmetry energy of finite nuclei at saturation density is often extracted by fitting ground state 
masses with various versions of the liquid drop mass formula. To this end, one needs to decompose the symmetry term of liquid 
drop into bulk (volume) and surface terms along the lines of the liquid droplet  model, and identify the volume symmetry energy 
coefficient as the symmetry energy derived from infinite nuclear matter at saturation density. Recently, there have been numerous 
efforts \cite{SHE07} to constrain the density dependence of the symmetry energy of infinite nuclear matter. Following the 
expression for the symmetry energy of finite nuclei at normal nuclear density by Danielewicz \cite{DAN03}, and using the constraint 
obtained from recent work on the the symmetry energy of infinite nuclear matter, one can empirically write the symmetry 
energy of a finite nucleus of mass $A$, as,

\begin{equation}
  S_{A}(\rho) = \frac{\alpha(\rho/\rho_{\circ})^{\gamma}}{1 + [\alpha(\rho/\rho_{\circ})^{\gamma}/\beta A^{1/3}]}
\end{equation}

where, $\alpha$ = 31 - 33 MeV, $\gamma$ = 0.55 - 0.69 and $\alpha/\beta$ = 2.6 - 3.0. The quantities $\alpha$ and $\beta$ are the 
volume and the surface symmetry energy at normal nuclear density. At present, the values of  $\alpha$, $\gamma$ and $\alpha/\beta$ 
remain unconstrained. The ratio of the volume symmetry energy to the surface symmetry energy ($\alpha/\beta$), is closely related 
to the neutron skin thickness \cite{DAN03}. Depending upon how the nuclear surface and the Coulomb contribution is treated, two 
different correlations between the volume and the surface symmetry energy have been predicted \cite{STE05} from fits to nuclear 
masses. Experimental masses and neutron skin thickness measurements for nuclei with $N/Z$ $>$ 1 should provide tighter constraint 
on the above parameters.

To compare the above empirical relation with the Thomas-Fermi calculation, we have assumed in Eq. 4 the same excitation 
energy dependence of the density of the expanding nucleus as obtained from the Thomas-Fermi calculation. The assumed excitation 
energy dependence of the density are shown by solid curves in the bottom most panel of Fig. 2 for nuclear masses of $A$ = 40, 
150 and 197. For comparison, we also show in this plot the densities obtained from the present study (solid circles and inverted 
triangles) and those from Ref. \cite{NATO02} (star symbols). The results of the empirical relation, Eq. 4, with $\alpha$ = 31.6 
MeV, $\gamma$ = 0.69 and $\alpha/\beta$ = 2.6, are shown by the dashed curves in top three panels of Fig. 2 for $A$ = 40, 150 
and 197. It is observed that the symmetry energy determined from the empirical relation (dashed curve) compares very well with 
the more formal Thomas-Fermi calculation (solid curve). The numerical values obtained from Eq. 4 agrees very well over a wide 
range of nuclear mass and excitation energy. Future measurements of symmetry energy as a function of excitation energy for very 
light and heavy nuclei should provide further insight into the validity and the theoretical understanding of the empirical relation 
and the Thomas-Fermi formalism.
 
In summary, the excitation energy dependence of the symmetry energy in multifragmentation reactions of 
$^{58}$Ni + $^{58}$Ni, $^{58}$Fe + $^{58}$Ni and $^{58}$Fe + $^{58}$Fe systems is studied. It is observed that the symmetry energy of a 
highly excited system decrease with increasing excitation energy.  The decrease is mainly due to the expansion that the system undergoes 
as its excitation energy increases. A comparison of the experimental data with the microscopic Thomas-Fermi calculation that accounts 
for the thermal and expansion effects in finite nuclei, shows good agreement. An empirical relation that can be used to study the 
symmetry energy of finite nuclei in various mass region is thereby proposed.
\  \\
\  \\
We thank Drs. J.N. De and S.K. Samaddar for making us available the results of their Thomas-Fermi calculation. We also thank 
Dr. J.B. Natowitz for providing us the data points from his studies on break-up density and caloric curve. This work was supported 
in part by the Robert A. Welch Foundation through grant No. A-1266, and the Department of Energy through grant No. DE-FG03-93ER40773.

% The Appendices part is started with the command \appendix;
% appendix sections are then done as normal sections
% \appendix

% \section{}
% \label{}

\newpage

\begin{figure}
\begin{center}
\resizebox{0.75\textwidth}{!}{
\includegraphics{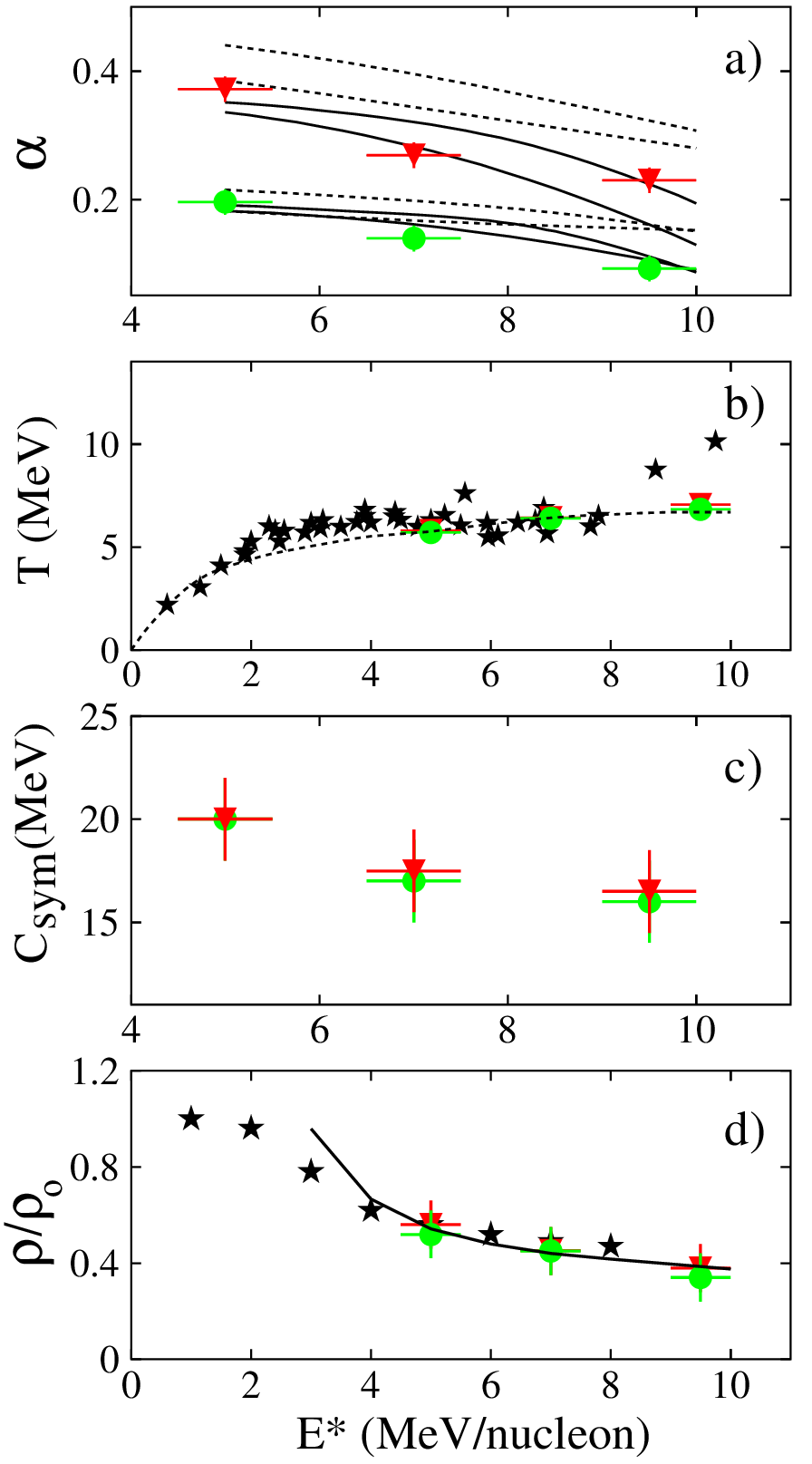}
}
%\vspace{5cm}  % Give the correct figure height in cm
\end{center}
\caption{(Color online) Isoscaling parameter $\alpha$, temperature T, symmetry energy $C_{sym}$, and density as a function of 
    excitation energy for the Fe + Fe and Ni + Ni pair of reaction (inverted triangles), and Fe + Ni and Ni + Ni pair of reactions 
    (solid circles) for the 30, 40 and 47 MeV/nucleon. a) Experimental isoscaling parameter as a function of excitation energy. The 
    solid and the dashed curves are the SMM calculations as discussed in the text. b) Temperature as a function of excitation energy. 
    The solid stars are taken from Ref. \cite{NAT02}. The dashed curve corresponds to the one determined from Eq. 3. c) Symmetry energy 
    as a function of excitation energy. d) Density as a function of excitation energy. The solid stars are from Ref. \cite{NATO02}. The 
    solid curve is from Ref. \cite{BON85}.}
\label{fig:1}       
\end{figure}

\begin{figure}
\begin{center}
\resizebox{0.85\textwidth}{!}{
\includegraphics{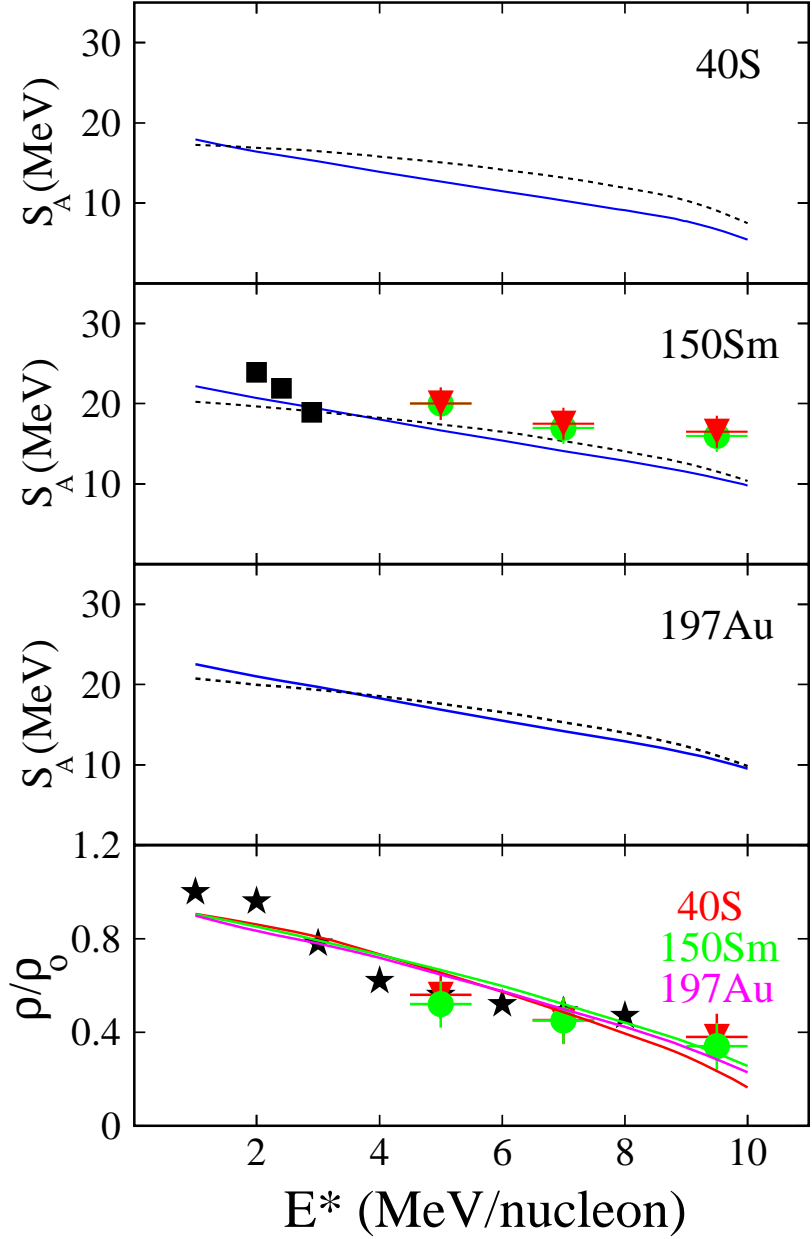}
}
%\vspace{5cm}       % Give the correct figure height in cm
\end{center}
\caption{(Color online) Top three panels: Symmetry energy as a function of excitation energy for $A$ = 40, 150 and 197. The solid 
blue curve corresponds to the Thomas-Fermi calculation and the dotted curve to the empirical relation Eq. 4 as discussed in the 
text. Bottom panel: Density as a function of excitation energy. The curves are from the Thomas-Fermi calculation. The star symbols are 
data from Ref. \cite{NAT02}. The solid circle and inverted triangle symbols are from the present study.}
\label{fig:5}       
\end{figure}

\end{document}